\documentclass[11pt]{article}
\usepackage{fullpage}
\usepackage{amsmath,amssymb}
\def\b{\backslash}
\def\tr{\textrm}

\def\mc{\mathcal}

\def\nbp{\tr{negbp}}
\def\pbp{\tr{posbp}}

\def\E{\mathrm{E}}
\def\P{\mathrm{Pr}}
\def\pd{\vec{p}}
\def\H{H}

\def\off{\textsf{off}}
\def\ee{(\epsilon^+,\epsilon^*,\epsilon^-)}
\def\ez{(\epsilon,\epsilon,0)}

\def\ps{\mathsf{qval}}

\newtheorem{thm}{Theorem}

\newtheorem{lemma}[thm]{Lemma}

\newtheorem{cor}[thm]{Corollary}

\begin{document}

\bibliographystyle{plain}

\title{Bloom maps} 
\author{David Talbot\thanks{Department of Informatics, University of Edinburgh, EH8 9LE, Scotland, UK. Email: d.r.talbot@sms.ed.ac.uk}\and John Talbot\thanks{
Department of Mathematics, University College London, WC1E 6BT, UK. 
Email: talbot@math.ucl.ac.uk.  }}
\date\today
\maketitle
\begin{abstract}

We consider the problem of succinctly encoding a static map to support approximate queries. We derive upper and lower bounds on the space requirements in terms of the error rate and the entropy of the distribution of values over keys: our bounds differ by a factor $\log e$. 

For the upper bound we introduce a novel data structure, the \emph{Bloom map}, generalising the Bloom filter to this problem. The lower bound follows from an information theoretic argument.

\end{abstract}
\section{Introduction}
The ability to query a map to retrieve a \emph{value} given a \emph{key} is fundamental in computer science. As the universe from which keys are drawn grows in size, information theoretic lower bounds imply that any data structure supporting error-free queries of a map requires unbounded space per key. However, if we are willing to accept errors, constant space per key is sufficient.

For example, in information retrieval we may wish to query the frequencies (values) of word sequences (keys) in documents. \emph{A priori} these sequences are drawn from a universe that is exponential in the length of a sequence. Returning an incorrect value for a small proportion of queries may be acceptable, if this enables us to support queries over a far larger data set.

Consider a map consisting of $n$ key/value pairs $M=\{(x_1,v(x_1)),(x_2,v(x_2)),\ldots,(x_n,v(x_n))\}$, where the keys $X=\{x_1,x_2,\ldots,x_n\}$ are drawn from a large universe $U$ and each value $v(x)$ is drawn from a fixed set of possible values $V=\{v_1,v_2,\ldots,v_b\}$. Suppose further that the distribution of values over keys is given by $\pd=(p_1,p_2,\ldots,p_b)$.  Thus if $X_i=\{x\in X\mid v(x)=v_i\}$ then $|X_i|=p_in$.
 
We consider the problem of constructing a space-efficient data structure supporting queries on $M$. For any key $x\in U$ the data structure should return the associated value $v(x)$ if $x\in X$, otherwise (i.e. if $x\in U\b X$) it should return $\bot\not\in V$. 

Using an information theoretic argument, we derive lower bounds on the space required to solve this problem when errors are allowed. These lower bounds are in terms of the error rate and the entropy of the distribution of values over keys $H(\pd)$.

We introduce the \emph{Bloom map}, a data structure generalising the Bloom filter \cite{Bloom70} to the approximate map problem. The space requirements of this data structure are within a $\log e$ factor of the lower bound. To be precise for an error rate of $\epsilon$ the Bloom map uses $\log e(\log 1/\epsilon + H(\pd))$ bits per key. 

To our knowledge, this paper is the first to make use of the distribution of values over keys to analyse the approximate map problem. Moreover, the Bloom map is the first data structure to take advantage of this distribution to save space by using variable length codes for distinct values. In the many practical settings where distributions with low entropy are encountered we expect the Bloom map to be of significant interest.

The main prior work on the approximate map problem is the Bloomier filter introduced by Chazelle et al.~\cite{Bloomier}. 
To store key/value pairs with values drawn from  a range of size $b$ with false positive probability $\epsilon$, the Bloomier filter requires $\alpha(\log 1/\epsilon + \log b)$ bits per key effectively using a fixed width encoding for any value in the range. It always returns the correct value for any $x\in X$. 

The Bloomier filter uses a perfect hash function introduced earlier by Czech et al.~\cite{Czech} whose analysis implies that the optimal $\alpha$ is approximately $1.23$. A simple calculation shows that in many cases the Bloom map will use less space. In fact it is also straightforward to extend the Bloom map to make use of the same family of perfect hash functions thereby reducing its space requirements to $1.23(\log1/\epsilon + H(\pd))$. 

In the next section we give a complete statement of the problem and prove lower bounds on the space requirements of any data structure supporting approximate queries of a static map with bounded errors (our most general result is Theorem \ref{lb:thm}). In section \ref{nbm:sec} we introduce the Simple Bloom map, a data structure supporting approximate queries that has near-optimal space requirements. In section \ref{bm:sec} we present more computationally efficient versions of the Bloom map.

\section{Problem statement and lower bounds}\label{lb:sec}

Consider a map of $n$ key/value pairs $M=\{(x_1,v(x_1)),(x_2,v(x_2)),\ldots,(x_n,v(x_n))\}$, where the keys $X=\{x_1,x_2,\ldots,x_n\}$ are drawn from a large universe $U$ of size $u$ and each value $v(x)$ is drawn from a fixed set of possible values $V=\{v_1,v_2,\ldots,v_b\}$. Suppose further that the distribution of values over keys is given by $\pd=(p_1,p_2,\ldots,p_b)$, where $\sum_{i=1}^bp_i=1$ and $\min_{i\in [b]}p_i>0$.  Thus if $X_i=\{x\in X\mid v(x)=v_i\}$ then $|X_i|=p_in$. We call such a collection $M$ of key/value pairs a \emph{$\pd$-map}.

We consider the problem of constructing a space-efficient data structure supporting queries on a static $\pd$-map $M$. For any key $x\in U$ the data structure should return the associated value $v(x)$ if $x\in X$, otherwise it should return $\bot\not\in V$. We will be interested in the case when $n$ is large, $u\gg n$ and $b,\pd=(p_1,p_2,\ldots,p_b)$ are constant. 

Given $u$, $n$, $b$ and $\pd=(p_1,p_2,\ldots,p_b)$ the total number of distinct $\pd$-maps is
\[
\binom{u}{n}\binom{n}{p_1n,p_2n,\ldots,p_bn}.\]
By Stirling's formula the multinomial coefficient is $2^{nH(\pd)+O(\log n)}$, where $\H(\pd)=-\sum_{i=1}^bp_i\log p_i$ is the entropy of $\pd$. (Logarithms here and elsewhere are base two.) Hence to distinguish between all $\pd$-maps without errors we require $m\geq n( \log u - \log n +\H(\pd)+o(1))$ bits.  For $n$ large and $u\gg n$ this is prohibitive: in particular we require more than a constant number of bits per key. Hence we are obliged to consider lossy data structures.

There are three distinct types of error that we will consider:
\emph{(False positives)}  $x\in U\b X$ is incorrectly assigned a value $v_i\in V$; \emph{(False negatives)} $x\in X_i$ is incorrectly assigned the value $\bot$; \emph{(Misassignments)} $x\in X_i$ is incorrectly assigned a value $v\in V\b \{v_i\}$.

Let $s$ be a binary string supporting queries by keys $x\in U$, i.e.~$s:U\to V\cup\{\bot\}$. Suppose that we use $s$ to encode a $\pd$-map $M$ with key set $X$. We wish to bound the proportion of keys on which $s$ returns an incorrect value. For $i\in [b]$ we define
\[
f^+(s)=\frac{|\{x\in U\b X\mid s(x)\neq \bot\}|}{|U\b X|},\]\[
 f^*_i(s)=\frac{|\{x\in X_i\mid s(x)\in V\b \{v_i\}\}|}{|X_i|},\qquad  f^-_i(s)=\frac{|\{x\in X_i\mid s(x)=\bot\}|}{|X_i|}.\]
Thus $f^+(s)$ is the proportion of false positives returned on $U\b X$, $f_i^*(s)$ is the proportion of misassigned values on $X_i$  and $f_i^-(s)$ is the proportion of false negatives returned on $X_i$. 

Given constants $\epsilon^+>0$ and $\epsilon^*,\epsilon^-\geq 0$ we will say that $s$ \emph{$\ee$-encodes} $M$ if it satisfies: $f^+(s)\leq \epsilon^+$ and, for all $i\in [b]$, $f_i^*(s)\leq \epsilon^*$ and $f_i^-(s)\leq \epsilon^-$. (We will assume throughout that $\max\{\epsilon^+,\epsilon^*,\epsilon^-\}<1/8$.)

If the only errors we allow are false positives then we have an $(\epsilon^+,0,0)$-encoding data structure.  (An example of such a data structure is the Bloomier filter \cite{Bloomier}). Theorem \ref{1lb:thm} gives lower bounds on the space requirements of such a data structure. (The proof  follows a counting argument generalising the argument applied to the approximate set membership problem by Carter et al.~\cite{carter}.)
\begin{thm}\label{1lb:thm}  The average number of bits required per key in any data structure that $(\epsilon^+,0,0)$-encodes all $\pd$-maps is at least $$\log 1/\epsilon^++\H(\pd)+o(1).$$
\end{thm}
\emph{Proof.}
Suppose that the $m$-bit string $s$ $(\epsilon^+,0,0)$-encodes some particular $\pd$-map $M$ with key set $X$. For $i\in [b]$ let $A_i^{(s)}=\{x\in U\mid s(x)=v_i\}$, $a_i^{(s)}=|A_i^{(s)}|$ and define $q_i^{(s)}$ by $a_i^{(s)}=p_in+\epsilon^+(u-n)q_i^{(s)}$. Since $X_i=\{x\in X\mid v(x)=v_i\}$ has size $p_in$ and $s$ always answers correctly on $X_i$ we have $q_i^{(s)}\geq 0$.

The proportion of $x\in U\b X$ for which $s(x)\neq \bot$ is $\sum_{i=1}^b \epsilon^+ q_i^{(s)}$. Since  $f^+(s)\leq \epsilon^+$, this implies that $\sum_{i=1}^b q_i^{(s)}\leq 1$.

If $N$ is any $\pd$-map with key set $Y$ that is also $(\epsilon^+,0,0)$-encoded by $s$ then, since $s$ correctly answers all queries on keys in $Y$, we have $Y_i=\{y\in Y\mid v(y)=v_i\}\subseteq A_i^{(s)}$, for all $i\in [b]$. Hence, since $|Y_i|=p_in$, $s$ can $(\epsilon^+,0,0)$-encode at most the following number of distinct $\pd$-maps
\[
\prod_{i=1}^b\binom{a_i^{(s)}}{p_in}=\prod_{i=1}^b\binom{p_in+\epsilon^+(u-n)q_i^{(s)}}{p_in}.\]
Choosing  $q_1,q_2,\ldots,q_b\geq 0$ to maximise this expression, subject to $\sum_{i=1}^b q_i\leq 1$, we have
\begin{equation*}\label{mem:eq}
2^m\prod_{i=1}^b\binom{p_in+\epsilon^+(u-n)q_i}{p_in}\geq \binom{u}{n}\binom{n}{p_1n,p_2n,\ldots,p_bn}.\end{equation*}
Using the fact that $\frac{(a-b)^b}{b!}\leq\binom{a}{b}\leq\frac{a^b}{b!}$ and taking logarithms we require
\[
m+\sum_{i=1}^bp_in\log(p_in +\epsilon^+(u-n) q_i)\geq n\log (u-n).\]
Dividing by $n$, recalling that $\sum_{i=1}^b p_i=1$ and rearranging we obtain
\[
\frac{m}{n}\geq \log 1/\epsilon^++\sum_{i=1}^bp_i\log 1/q_i-\sum_{i=1}^b p_i \log \left(1+\frac{n(p_i-\epsilon^+q_i)}{\epsilon^+q_iu}\right)+\log\left(1-\frac{n}{u}\right).\]
Our assumption that $u\gg n$ (which is equivalent to $n/u=o(1)$) together with the fact that $\log (1+\alpha)=O(\alpha)$ for $\alpha$ small implies that the last two terms are $o(1)$. Hence the average number of bits required per key satisfies
\[
\frac{m}{n}\geq\log1/\epsilon^++\sum_{i=1}^bp_i\log 1/q_i +o(1).\]
Gibbs' inequality implies that the sum is minimised when $q_i=p_i$ for all $i\in [b]$, the result follows.$\hfill\Box$

\medskip
This calculation can be extended to the case when errors are also allowed on keys in the set $X$. 
\begin{thm}\label{lb:thm}  The average number of bits required per key in any data structure that $\ee$-encodes all $\pd$-maps is at least \[
(1-\epsilon^-)\log1/\epsilon^+ +(1-\epsilon^- -\epsilon^*)\H(\pd)-\H(\epsilon^-,\epsilon^*,1-\epsilon^--\epsilon^*)+o(1).\]
\end{thm}
\emph{Proof:} The basic idea behind the proof of this result is the same as that of Theorem \ref{1lb:thm}, however the details are somewhat more involved. (See Appendix.) $\hfill\Box$

The Bloom map, which we introduce in Section \ref{nbm:sec}, is $\ez$-encoding. To enable us to evaluate how far its space requirements are from optimal we give the following simple corollary. 
\begin{cor}\label{bmlb:thm}  The average number of bits required per key in any data structure that $(\epsilon,\epsilon,0)$-encodes all $\pd$-maps is at least \[(1-\epsilon)(\log1/\epsilon +\H(\pd)-(\epsilon+\epsilon^2))+o(1).\]
\end{cor}
\emph{Proof:} Substitute $\epsilon^+=\epsilon^*=\epsilon$ and $\epsilon^-=0$ into Theorem \ref{lb:thm} and use $\log(1-\epsilon)\geq -(\epsilon+\epsilon^2)$.$\hfill \Box$

 \section{The Simple Bloom map}\label{nbm:sec}
Let $M$ be a $\pd$-map with key set $X$. Thus, for $i\in [b]$, $X_i=\{x\in X\mid v(x)=v_i\}$ has size $p_in$. Our first succinct data structure supporting queries for $M$ is the Simple Bloom map. This is constructed by simply storing the values directly in a Bloom filter. 

Let $B$ be an array of size $m$ that is initially empty. For each $i\in [b]$ we choose $k_i\geq 1$ independent random hash functions $h_{i,j}:U\to [m]$ (we will explain how to set $k_1,k_2,\ldots,k_b$ optimally below). To store the key/value pair $(x,v_i)$ we compute $h_{i,j}(x)$ for each $j\in [k_i]$ and set the bits $B[h_{i,j}(x)]$ to one. To query $B$ with a key $x\in U$ we compute $h_{i,j}(x)$ for each $i\in [b], j\in [k_i]$ and set \[
\ps(x)=\left\{i\in [b]\mid \wedge_{j=1}^{k_i}B[h_{i,j}(x)]=1\right\}.\]
If $\ps(x)=\emptyset$ we return $\bot$ otherwise we return $v_c$, where $c=\max \ps(x)$. Note that if $(x,v_i)\in M$ then $i\in \ps(x)$ and so $\bot$ is never returned when querying $x$, i.e.~there are no false negatives. However both false positives and misassignments can occur. 

Let $t=n\sum_{i=1}^b p_ik_i$ be the total number of hashes performed during the creation of $B$. Let $\rho$ be the proportion of bits that remain zero in $B$. If $f^+(B)$ is the false positive probability of $B$, i.e.~the probability that $B$ returns $v\neq \bot$ for a fixed $x\in U\b X$, then\[
f^+(B)=\P\{\ps(x)\neq \emptyset\}\leq \sum_{i=1}^b \P\{i\in \ps(x)\}=\sum_{i=1}^b (1-\rho)^{k_i}.\]
If $f^*_i(B)$ is the misassignment probability for $B$ over keys in $X_i$, i.e.~the probability that $B$ returns $v\in V\b\{v_i\}$ for a fixed $x\in X_i$, then 
\[
f_i^*(B)=\P\{\max \ps(x)>i\}\leq \sum_{j=i+1}^b \P\{j\in \ps(x)\}=\sum_{j=i+1}^b (1-\rho)^{k_j}<\sum_{i=1}^b(1-\rho)^{k_i}.\]
Hence in order to minimise $f^+(B)$ and $f^*_i(B)$ we consider the constrained optimisation problem: minimise $\sum_{i=1}^b (1-\rho)^{k_i}$ subject to $\sum_{i=1}^bp_ik_i=t/n$.
A standard application of Lagrange multipliers yields the solution
\[
k_i=\frac{t}{n}+\frac{H(\pd)+\log p_i}{\log(1-\rho)}.\]
For this choice of the $k_i$ we have
\[
 \sum_{i=1}^b (1-\rho)^{k_i}=(1-\rho)^{t/n}2^{H(\pd)}\sum_{i=1}^bp_i=2^{H(\pd)}(1-\rho)^{t/n}.\]
By a simple martingale argument, identical to that given by Mitzenmacher  \cite{MitzenmacherCBF} for the Bloom filter, $\rho$ is extremely close to its expected value if $t=O(m)$ (see Appendix). Assuming $\rho\geq \E[\rho]$ we have
\[
 2^{H(\pd)}(1-\rho)^{t/n}\leq 2^{H(\pd)}\left(1-\left(1-\frac{1}{m}\right)^t\right)^{t/n}.\]
This last expression (without the factor $2^{H(\pd)}$) is familiar from the standard Bloom filter error analysis: it is minimised at $t=m\ln 2$, when it equals $2^{H(\pd)-\frac{m}{n}\ln2}$. (Note that as for the standard Bloom filter the expected proportion of bits set in $B$ is $1/2$.)

 Thus to guarantee $f^+(B)\leq \epsilon$ and $f_i^*(B)\leq \epsilon$ for all $i\in [b]$ it is sufficient to take 
\[
m=n\log e(\log1/\epsilon+H(\pd)),\qquad 
k_i=\log1/\epsilon+\log1/p_i \quad\tr{for $i\in[b]$}. 
\]
(As with the standard Bloom filter, the $k_i$ must be integers, for simplicity we will ignore this.)

Since Corollary \ref{bmlb:thm} gives a lower bound for the space required by an $\ez$-encoding data structure we would like to claim that $B$ is $\ez$-encoding. This is not quite true: the \emph{expected} proportion of false positives and misassignments is at most $\epsilon$ but this does not guarantee that $B$ is $\ez$-encoding. However $B$ is still essentially $\ez$-encoding since, with high probability, the proportion of false positives or misassignments is at most $\epsilon+O(1/\sqrt{n})$. (See Appendix for details.)
\begin{thm}\label{nbm:thm}
The Simple Bloom map $\ez$-encodes all $\pd$-maps and uses $\log e(\log 1/\epsilon +H(\pd))$ bits per key. 
\end{thm}
Note that by Corollary \ref{bmlb:thm} the space requirements of the Simple Bloom map are essentially a factor $(1-\epsilon)^{-1}\log e$ from optimal, for $\epsilon\leq 0.01$  this is less than $1.46$.

We remark that an $(\epsilon,\epsilon,\epsilon)$-encoding data structure can be created from the Simple Bloom map by simply discarding $\epsilon p_in$ keys from $X_i$ for each $i\in [b]$. The amount of memory saved is $\epsilon n\log e (\log 1/\epsilon +H(\pd))$ (cf.~Theorem \ref{lb:thm}).

Although the Simple Bloom map is succinct it suffers from two obvious drawbacks if $b$ is not small: the number of hashes/bit probes performed during a query and the number of independent hash functions required is $O(b\log (b/\epsilon))$. In section \ref{bm:sec} we explain how to overcome these problems by ``reusing'' hash functions and using an optimal binary search tree.
\section{Efficient Bloom maps}\label{bm:sec}
Let $M$ be a $\pd$-map that we wish to store.  Sort the list of probabilities of keys so that $p_1\geq p_2\geq \cdots \geq p_b$. Construct an optimal alphabetic binary tree $T(\pd)$ for $\pd$ with leaves labelled $v_1,v_2,\ldots,v_b$ (by for example the Garsia--Wachs algorithm, see Knuth \cite{knuth3} page 446). The label of a leaf $w$ is denoted by $\textsf{val}(w)$. Note that $T(\pd)$ is a full binary tree, i.e.~every node is either a leaf or has exactly two children.

For any binary tree $T$ let $r(T)$ denote its root and $T_L$, $T_R$ denote its left and right subtrees respectively. For any  node $w$ let $P_{w}$ denote the set of nodes on the path in $T$ from the root to $w$ and let $l_w=|P_w|-1$ be the depth of $w$.  For $d \geq 0$ let $T_d$ be the set of nodes in $T$ at depth $d$. 

We number the nodes in $T(\pd)$ from left to right at each level, starting at the root and going down. We call these numbers \emph{offsets}. (So the root has offset 0, its left child has offset 1 and its right child has offset 2 etc.) Note that all nodes have distinct offsets. The offset of a node  $w$ is denoted $\off(w)$. 
To each node $w\in T(\pd)$ we also associate an integer $k_w$. We will specify choices for the $k_w$ later, we first impose two simple conditions: $k_w\geq 1$ for all nodes and $k_w\geq \log1/\epsilon$ for all leaves. Set 
\begin{equation}\label{mkbm:eq}
m=\log e\left( \sum_{i=1}^bp_in\sum_{w\in P_{v_i}}k_w\right),\qquad k=\max_{i\in [b]}\sum_{w\in P_{v_i}}k_w.\end{equation}
We now impose a third condition on the $k_w$: they are chosen so that $m\leq 2n \log e\log (b/\epsilon)$.

Let $h_1,h_2,\ldots,h_k$ be independent random hash functions, $h_j:U\to [m]$. (We will refer to these as the \emph{base} hash functions.)  For a node $w$ let $s_w=\sum_{u\in P_w\b\{w\}}k_w$. We associate $k_w$ hash functions with $w$: $h_{w,1},h_{w,2},\ldots,h_{w,k_w}$, where $h_{w,j}:U\to [m]$ is defined by $h_{w,j}(x)=h_{s_w+j}(x)+\off(w) \mod m$.

The Bloom map $B$ is an array of size $m$ that is initially empty (all bits are zero).  To store a key/value pair $(x,v_i)$ we use the algorithm \textsf{Store}$(x,v_i,T(\pd),B)$ (see Figure \ref{bm:fig}). This does the following: for each node $w$ in the path $P_{v_i}$, starting from the root, it evaluates the associated $k_w$ hashes at $x$ and sets the corresponding bits in $B$. Note that (ignoring offsets) the hash functions used while storing $(x,v_i)$ are $h_1,h_2,\ldots,h_{t_i}$, where $t_i=\sum_{w\in P_{v_i}}k_w$. Hence the bits which are set in $B$ by \textsf{Store}$(x,v_i,T(\pd),B)$ are chosen independently and uniformly at random. Moreover, since each key is stored with at most one value, the entire process of storing the $\pd$-map in $B$ is equivalent to setting $t=n\sum_{i=1}^bp_it_i$ independently chosen random bits in $B$.
\begin{figure}
\begin{tabbing}
\sf{St}\=\sf{or}\=\sf{e}$($\=$x,v_i,T,B)$\hspace{2cm}\=\textsf{Qu}\=\textsf{ery}$(x,T,B)$\hspace{2cm}\=\sf{Fin}\=\sf{dv}\=\sf{al}$(x,v,T,B)$ \\
for $d=0$ to $l_{v_i}$\>\>\>\>$v\leftarrow \bot$ \>\>$w\leftarrow r(T)$\\
\>$w\leftarrow P_{v_i}\cap T_d$\>\>\>\textsf{Findval}$(x,v,T,B)$ \>\>for $j=1$ to $k_w$\\
\>for $j=1$ to $k_w$\>\>\> return $v$\>\>\>if $B[h_{w,j}(x)]=0$ then return false\\ 
\>\>$B[h_{w,j}(x)]\leftarrow 1$\>\>\>\> if $w$ is a leaf then\\
\>\>\>\>\>\>\>$v\leftarrow \textsf{val}(w)$; return true\\ 
\>\>\>\>\>\>if \textsf{Findval}$(x,v,T_R,B)$ then return true\\
\>\>\>\>\>\>return \textsf{Findval}$(x,v,T_L,B)$
\end{tabbing}
\caption{Storing and querying keys in a Bloom map} \label{bm:fig}
\end{figure}

To query $B$ with a key $x\in U$ we use the algorithm \textsf{Query}$(x,T(\pd),B)$. This calls \textsf{Findval}$(x,v,T(\pd),B)$ with $v$ initialised to $\bot$ and returns the value of $v$ when \textsf{Findval} $(x,v,T(\pd),B)$ terminates (see Figure \ref{bm:fig}). Starting with $T(\pd)$, \textsf{Findval} evaluates the hash functions associated with the root of the current tree, returning false if it finds a zero bit in $B$, otherwise it continues down the tree, first looking at the right subtree and then, if this fails, looking at the left subtree. If it reaches a leaf at which the corresponding bits in $B$ are all set then $v$ is assigned the value associated with this leaf and it returns true, otherwise the value of $v$ will remain equal to $\bot$.


By our choice of $m=t\log e$ the expected proportion of bits that remain zero in $B$ (once we have stored the $\pd$-map $M$) is $1/2$ and with high probability the actual proportion, which we denote by $\rho$, is very close to this. For simplicity we will assume that $\rho\geq 1/2$. 

We now consider the probability of errors. To simplify our analysis we assume that any leaf $v_i$ is at depth $\log1/p_i$ (since $T(\pd)$ is an optimal  alphabetic binary tree this is almost true). For $x\in U$ and $i\in [b]$ define 
\[\mc{H}_i(x)=\{h_{w,l}(x)\mid w\in P_{v_i},l\in [k_w]\},\qquad \ps(x)=\left\{i\in [b]\mid \wedge_{h\in\mc{H}_i(x)}B[h]=1\right\}.\] Thus $i\in \ps(x)$ iff all of the bits in $B$ indexed by the hash functions on the path $P_{v_i}$ evaluated at $x$ are set. If $\ps(x)=\emptyset$ then \textsf{Query} returns $\bot$, otherwise, since \textsf{Findval} always explores right subtrees first, it returns $v_c$, where $c=\max\ps(x)$. If $x\in X_i$ then $i\in \ps(x)$ and so no false negatives can occur. False positives and misassignments are possible, we consider the case of false positives first.

If $x\in U\b X$ then for fixed $i\in [b]$ the bits in $\mc{H}_i(x)$ are simply independent random choices from $[m]$. This is because  if $t_i=\sum_{w\in P_{v_i}}k_w$ then the hash functions we evaluate are simply offsets, modulo $m$, of the first $t_i$ of our base hash functions. By our assumptions that:  $k_w\geq 1$ for all nodes; $k_{v_i}\geq \log 1/\epsilon$ and $v_i$ is at depth $\log 1/p_i$, we have $t_i\geq -\log \epsilon p_i$. Since $\rho\geq1/2$ 
the false positive probability satisfies \[
f^+(B)=\P\{\ps(x)\neq \emptyset\}\leq\sum_{i=1}^b\P\{i\in \ps(x)\}\leq\sum_{i=1}^b(1-\rho)^{t_i}\leq\sum_{i=1}^b\frac{1}{2^{t_i}}\leq \epsilon.\]

Calculating the probability of a misassignment when $B$ is queried with $x\in X_i$ is more involved. Note that if an incorrect value $v_j\neq v_i$ is returned for $x\in X_i$ then $j>i$. For $i< j $ and $x\in X_i$ let $P_{i,j}=P_{v_j}\b P_{v_i}$ be the part of the path $P_{v_j}$ that is disjoint from the path  $P_{v_i}$ and  let $\mc{H}_{i,j}(x)=\{h_{w,l}(x)\mid w\in P_{i,j},l\in [k_w]\}$. The misassignment probability satisfies
\begin{equation}\label{mp:eq}
f^*_i(B)=\P\{\max \ps(x)>i\}\leq \sum_{j=i+1}^b \P\{j\in \ps(x)\}=\sum_{j=i+1}^b\P\left\{\wedge_{h\in \mc{H}_{i,j}(x)}B[h]=1\right\}.\end{equation}

To bound this probability we consider the following:
suppose that rather than storing all of the key/value pairs from $M$ in $B$ we had instead stored all of them \emph{except} $(x,v_i)$. Let $B'$ denote the resulting $m$-bit array.
 Let $t_{i,j}=|\mc{H}_{i,j}(x)|$. Since $(x,v_i)$ has not been stored in $B'$ we have (by the same argument as used for $f^+(B)$) that
\begin{equation}\label{mp2:eq}
\P\left\{\wedge_{h\in \mc{H}_{i,j}(x)}B'[h]=1\right\}\leq \frac{1}{2^{t_{i,j}}}.\end{equation}

If all of the bits in $B$ indexed by elements in $\mc{H}_{i,j}$ are set then either they are all set in $B'$ or there must be at least one bit in $\mc{H}_{i,j}$ that is only set  once $(x,v_i)$ is stored. The later case can only occur if $\mc{H}_i\cap \mc{H}_{i,j}\neq \emptyset$. Hence
\begin{equation}\label{mp3:eq}
\P\left\{\wedge_{h\in \mc{H}_{i,j}(x)}B[h]=1\right\}\leq \P\left\{\wedge_{h\in \mc{H}_{i,j}(x)}B'[h]=1\right\}+\P\{\mc{H}_i\cap \mc{H}_{i,j} \neq \emptyset\}.\end{equation}
If $\hat{h}_1\in \mc{H}_i(x)$ and $\hat{h}_2\in \mc{H}_{i,j}(x)$ then $\P\{\hat{h}_1=\hat{h}_2\}$ is either $1/m$ or $0$, since $\hat{h}_1$ and $\hat{h}_2$ either use different base hash functions (and so are independent and random in $[m]$) or they use the same base hash function with different offsets and hence are distinct.

Recall that $k=\max_{i\in[b]}\sum_{w\in P_{v_i}}k_w$. If $c\in [b]$ satisfies $k=\sum_{w\in P_{v_c}}k_w$ then $m\geq np_ck\log e $. Moreover the $k_w$ were chosen so that $n\leq m\leq 2n \log e\log (b/\epsilon)$. Hence 
\begin{equation}\label{mp4:eq}
\P\{\mc{H}_i\cap \mc{H}_{i,j}\neq \emptyset\}\leq \frac{|\mc{H}_i|\cdot |\mc{H}_{i,j}|}{m}\leq \frac{k^2}{m}\leq  \left(\frac{m}{np_c\log e}\right)^2\frac{1}{m}\leq \left(\frac{2\log b/\epsilon}{p_c}\right)^2\frac{1}{n}=O\left(\frac{1}{n}\right),\end{equation}
 where the final equality uses our assumption that $\pd$, $b$ and $\epsilon$ are constant.

Combining (\ref{mp:eq}), (\ref{mp2:eq}), (\ref{mp3:eq}) and (\ref{mp4:eq}) we obtain
\begin{equation}\label{mp5:eq}
f^*_i(B)\leq \sum_{j=i+1}^b \frac{1}{2^{t_{i,j}}}+O\left(\frac{1}{n}\right)\approx \sum_{j=i+1}^b\frac{1}{2^{t_{i,j}}},\end{equation}
where $t_{i,j}=\sum_{w\in P_{i,j}}k_w$. Thus to ensure $f^*_i(B)\leq \epsilon$ we choose the $k_w$ so that $\sum_{j=i+1}^b2^{-t_{i,j}}\leq \epsilon$. There are various ways in which this can be done and exactly how we choose the $k_w$ will effect not only $f_i^*(B)$ but also the memory required to store the Bloom map and the amount of work we expect to do when querying it. Since different space/time trade-offs may be of interest in different applications we define two special types of Bloom map: Standard and Fast.
\begin{itemize}
\item\emph{(Standard)} $k_w=1$ for all internal nodes (i.e.~all non-leaf nodes), $k_{v_i}= \log1/\epsilon + \log (H_b -1)+1$ for all leaves (where $H_{b}=\sum_{l=1}^b 1/l$ is the $b$th Harmonic number).
\item\emph{(Fast)} $k_w=2$ for all internal nodes, $k_{v_i}=\log 1/\epsilon+2$ for all leaves.
\end{itemize}
\begin{thm}\label{3bm:thm}
The Standard and Fast Bloom maps are both $\ez$-encoding for $\pd$-maps. The average number of bits required per key is:
\begin{itemize}
\item (Standard): $\log e(\log 1/\epsilon+ H(\pd)+\log (H_b-1)+1)$.
\item (Fast): $\log e(\log 1/\epsilon+ 2H(\pd)+2)$.
\end{itemize}

If $x\in U\b X$ then the expected number of bit probes performed during \textsf{Query}$(x,T(\pd),B)$ is at most: (Standard) $H(\pd)+2$; (Fast) $3$.

If $x\in X_i$ then the expected number of bit probes performed during \textsf{Query}$(x,T(\pd),B)$  is at most:
(Standard) $O((\log b)^2)+\log 1/p_i+\log1/\epsilon$;  (Fast) $3\log (b-i+1)+2\log 1/p_i+\log1/\epsilon+2$.
\end{thm} 
The Standard Bloom map uses little more than a factor $(1-\epsilon)^{-1}\log e$ extra bits per key than the lower bound of Corollary \ref{bmlb:thm}. (In addition to the factor of $(1-\epsilon)^{-1}\log e$ it uses at most an extra $1+\log\log b$ bits per key, since $H_b<\log b$.)  The Fast Bloom map uses slightly more space but has the advantage of using significantly fewer bit probes when querying keys: in particular we expect to perform at most 3 bit probes on $x\in U\b X$. In any case the Fast Bloom map uses less than 2.9 times as much memory per key as the lower bound and if $H(\pd)$ is small compared to $\log 1/\epsilon$ this factor will be much closer to 1.46.

We note that other choices for the $k_w$ are possible and depending on the application may be desirable. For example, altering the Fast Bloom map by adding $s\geq 1$ to $k_r$, where $r$ is the root of $T(\pd)$, yields a Bloom map that will perform $2+1/2^s$ bit probes on average, for $x\in U\b X$. Another possibility is to alter the Standard Bloom map by adding $\log(H(\pd)+2)$ to the value of $k_r$ giving a Bloom map which performs the same expected number of bit probes as the Fast Bloom map on $x\in U\b X$ and the same expected number of bit probes as the Standard Bloom map on $x\in X$.

\emph{Proof of Theorem \ref{3bm:thm}:}
We first show that both Bloom maps are $\ez$-encoding. We know already that $f^+(B)\leq \epsilon$ so we consider $f^*_i(B)$. We require the following simple lemma.
\begin{lemma}\label{leaf:lem}
Let $T$ be a full binary tree with leaves $v_1,v_2,\ldots, v_b$ at depths $l_1\leq l_2\leq \cdots\leq l_b$.
\begin{itemize} 
\item[(a)] If $1\leq i < j \leq b$ then the number of nodes in $P_{v_j}\b P_{v_i}$ is at least $\log \left(\sum_{k=i}^j2^{l_j-l_k}\right)$.
\item[(b)] If $T_d$ is the set of nodes in $T$ at depth $d$ then
\[
\sum_{d=0}^{l_b}\frac{|T_d|}{2^d}\leq1+\sum_{i=1}^b \frac{l_i}{2^{l_i}}.\]
\item[(c)] The number of left branches on the path $P_{v_i}$ is at most $\log(b-i+1)$. 
\end{itemize}
\end{lemma}
\emph{Proof:} These are all straightforward, see Appendix for details. $\hfill\Box$

Lemma \ref{leaf:lem} (a), together with our assumption that $v_k$  is at depth $\log 1/p_k$ in  $T(\pd)$ and the fact that $p_1\geq p_2\geq \cdots \geq p_b$ implies that the number of internal nodes on $P_{i,j}$ is at least $\log\left(\sum_{j=i}^k p_k/p_j\right)-1$. Let $a$ be the common value of $k_w$ for all internal nodes. By (\ref{mp5:eq}) we have
\[
f^*_i(B)\leq \sum_{j=i+1}^b \frac{1}{2^{t_{i,j}}}\leq \sum_{j=i+1}^b\left(\frac{p_j}{\sum_{k=i}^j p_k}\right)^a \frac{1}{2^{k_{v_j}-a}}\leq \sum_{j=i+1}^b\frac{1}{(j-i+1)^a2^{k_{v_j}-a}},\]
where the last inequality follows from the fact that $p_j\leq p_k$ for all $i\leq k \leq j$. In the case of the Standard Bloom map we have $a=1$ and $k_{v_j}=\log 1/\epsilon+\log (H_{b}-1)+1$, hence $f^*_i(B)\leq \epsilon$. For the Fast Bloom map $a=2$, $k_{v_j}=\log1/\epsilon+2$ and $\sum_{k=1}^\infty 1/l^{2}=\pi^2/6$ imply that
\[
f_i^*(B)\leq \sum_{l=2}^{b-i+1}\frac{\epsilon}{l^2}\leq \epsilon\left(\frac{\pi^2}{6}-1\right)< \epsilon.\]
Hence both Bloom maps are $\ez$-encoding.

 Now consider how much work we expect to do when querying $B$. We measure this in terms of the expected number of bit probes performed. (Note that as described each bit probe performed by \textsf{Findval} involves the evaluation of a hash function, this need not be the case. The use of offsets ensures that we never need to evaluate more than $k=\max_{i\in [b]}\sum_{i\in P_{v_i}}k_w$ base hash functions, different offsets can then be added as required.) We consider the cases $x\in X$, $x\in U\b X$ separately.

Let $\nbp$ denote the expected number of bit probes performed by \textsf{Query}$(x,T(\pd),B)$ for $x\in U\b X$. 
The easiest case is the Fast Bloom map, in which every internal node $w$ has $k_w=2$. Let $\nbp(T)$ be the expected number of bit probes performed by \textsf{Findval} in a tree $T$. We wish to find $\nbp=\nbp(T(\pd))$. Starting from the root of $T(\pd))$ we have 
\[
\nbp(T(\pd))\leq 1+\frac{1}{2}+\frac{1}{4}(\nbp(T_L(\pd))+\nbp(T_R(\pd))),\]
since if $b_1,b_2$ are the first two bit probes then $\P\{b_1=0\}=\rho\geq 1/2$ and $\P\{b_1=b_2=1\}=(1-\rho)^2\leq 1/4$.
Iterating and using the fact that all nodes in $T(\pd)$ have at least two associated bit probes we find 
\[
\nbp\leq  \frac{3}{2}\sum_{j=0}^\infty\frac{1}{2^j}=3.\]

In the Standard Bloom map $k_w=1$ for every internal node,  hence if $w$ is at depth $l_w$ then the probability that the bit probe associated with $w$ is evaluated during \textsf{Query}$(x,T(\pd),B)$,  is at most $2^{-l_w}$. Moreover for a leaf $v_i$ at depth $\log 1/p_i$ the probability that \textsf{Findval} performs more than one bit probe at $v_i$ is at most $p_i/2$ and in this case we expect to perform at most two extra bit probes at the leaf. Hence if $T_d(\pd)$ is the set of nodes in $T(\pd)$ at depth $d$ then the expected number of bit probes performed during \textsf{Query}$(x,T(\pd),B)$ is at most
\[
\nbp\leq 2\sum_{i=1}^b\frac{p_i}{2}+\sum_{d=0}^{\infty}\frac{|T_d(\pd)|}{2^d}=1+\sum_{d=0}^{\infty}\frac{|T_d(\pd)|}{2^d}.\]
By Lemma \ref{leaf:lem} (b) this is at most $H(\pd)+2$.

Finally we calculate the expected number of bit probes performed by \textsf{Query}$(x,T(\pd),B)$, for $x\in X_i$, which we denote by $\pbp(i)$.  This will be the number of bits set during \textsf{Store}$(x,v_i,T(\pd),B)$, plus the expected number of bit probes performed by \textsf{Findval} in the ``false subtrees'' it explores, where a false subtree is any maximal subtree disjoint from the path $P_{v_i}$. The number of false subtrees is simply the number of left branches in the path $P_{v_i}$, since at each such branch \textsf{Findval} first explores the right (false) subtree. By Lemma \ref{leaf:lem} (c) the number of false subtrees is at most $\log(b-i+1)$. To simplify our analysis we will assume that the bit probes in false subtrees are independent and random. By a similar argument to that used during the calculation of the misassignment probability above this is essentially true.

For the Fast Bloom map we expect to perform at most three bit probes in each false subtree. Since the number of false subtrees in $T(\pd)$ is at most $\log (b-i+1)$ the expected number of bit probes performed in false subtrees is at most $3\log (b-i+1)$. Since the number of bits set by \textsf{Store}$(x,v_i,T(\pd),B)$ is $2\log 1/p_i+\log 1/\epsilon +2$ we have \[
\pbp(i)\leq 3\log (b-i+1) +2\log 1/p_i+\log 1/\epsilon+2.\]

Now consider the Standard Bloom map. Any false subtree is a full binary tree with $z\leq b-i$ leaves and hence corresponds to an optimal binary search tree for some probability distribution $q=(q_1,q_2,\ldots,q_z)$. Since $H(q)\leq \log z \leq \log (b-i)$ the expected number of bit probes performed in any false subtree is at most $\log (b-i)+2$. The number of bits set by \textsf{Store}$(x,v_i,T(\pd),B)$ is $\log 1/p_i+\log 1/\epsilon+\log (H_b-1) + 1$. Hence
\[
\pbp(i)= O((\log b)^2)+\log1/p_i+\log 1/\epsilon.\]
This completes the proof of Theorem \ref{3bm:thm}.$\hfill\Box$ 
\bibliography{bloommap}
\section*{Appendix}
\emph{Proof of Theorem \ref{lb:thm}.}
Let $M$ be a fixed $\pd$-map with key set $X$. Suppose that $M$ is $\ee$-encoded by the $m$-bit string $s$. For $i \in [b]$ let $a_i^{(s)}=|\{x\in U\mid s(x)=v_i\}|$ and let $w_i^{(s)}=|\{x\in U\b X\mid s(x)=v_i\}|$. Define $q_i^{(s)}\geq 0$ by $w_i^{(s)}=\epsilon^+(u-n)q_i^{(s)}$. So $a_i^{(s)}\leq n+\epsilon^+(u-n)q_i^{(s)}$. Since $f^+(s)\leq \epsilon^+$ we have \begin{equation}\label{q:eq}\sum_{i=1}^b q_i^{(s)}\leq 1.\end{equation}

We now need to consider how many distinct $\pd$-maps $N=\{(y_1,v(y_1),(y_2,v(y_2),\ldots,(y_n,v(y_n))\}$ can be $\ee$-encoded by the string $s$. Let $Y$ be the key set of $N$ and for $i\in [b]$ let $Y_i=\{y\in Y\mid v(y)=v_i\}$, so $|Y_i|=p_in$. For $0\leq j \leq b$ let $y_{i,j}=|\{y\in Y_i\mid s(y)=v_j\}|$. 

Since $f_i^-(s)\leq \epsilon^-$,  $f_i^*(s)\leq \epsilon^*$ and  $s$ returns a value from $V\cup\{\bot\}$ for each element in $Y_i$  we have the following three constraints on the $y_{i,j}$
 \begin{equation}\label{y:eq}
y_{i,0}\leq \epsilon^- p_in,\qquad \sum_{j\in [b]\b \{i\}}y_{i,j}\leq \epsilon^*p_in, \qquad \sum_{j=0}^by_{i,j}=p_in.\end{equation}
We can now bound the number of choices for the $y_{i,j}$. Since $\sum_{j=0}^b y_{i,j}=p_in$,  $y_{i,i}$ is determined by fixing the values of $y_{i,j}$ for $j \neq i$. Hence the number of choices for the $y_{i,j}$ is at most
\[
\epsilon^-p_in\sum_{l=0}^{\lfloor \epsilon^* p_i n\rfloor}\binom{l+b-2}{b-2}\leq \epsilon^-\epsilon^* (p_i n)^2\binom{\epsilon^* p_i n +b-2}{b-2}.\](This is because (\ref{y:eq}) implies that there are at most $\epsilon^-p_in$ choices for $y_{i,0}$ while $\sum_{j\in [b]\b\{i\}}y_{i,j}=l$ for some integer $0\leq l\leq \epsilon^* p_in$. The number of ways of choosing $b-1$ non-negative integers whose sum is $l$ is $\binom{l+b-2}{b-2}$.)
For a particular choice of the $y_{i,j}$ the number of choices for the keys in $Y_i$ is at most
\begin{equation}\label{yi:eq}
\binom{u}{y_{i,0}}\prod_{j=1}^b\binom{n+\epsilon^+(u-n)q_j^{(s)}}{y_{i,j}}.\end{equation}
(This is because any particular choice for the keys in $Y_i$ is given by choosing $y_{i,0}$ keys on which $s$ returns $\bot$ and then choosing $y_{i,j}$ keys on which $s$ returns $v_j$, for each $j\in [b]$.)

Let $y'_{i,0},y'_{i,1},\ldots y'_{i,b}$ be chosen to maximise (\ref{yi:eq}) subject to $(\ref{y:eq})$. The number of choices for the keys in $Y_i$ is at most
\[
\epsilon^-\epsilon^* (p_i n)^2\binom{\epsilon^* p_i n +b-2}{b-2}\binom{u}{y'_{i,0}}\prod_{j=1}^b\binom{n+\epsilon^+(u-n)q_j^{(s)}}{y'_{i,j}}.\]
Hence the total number of $\pd$-maps which can be $\ee$-encoded by the string $s$ is at most
\[
\prod_{i=1}^b\left(\epsilon^-\epsilon^* (p_i n)^2\binom{\epsilon^* p_i n +b-2}{b-2}\binom{u}{y'_{i,0}}\prod_{j=1}^b\binom{n+\epsilon^+(u-n)q_j^{(s)}}{y'_{i,j}}\right).\]
Letting $q_1,q_2,\ldots,q_b\geq 0$ be chosen to maximise this expression subject to $\sum_{j=1}^bq_j\leq 1$ we obtain
\begin{multline*}
2^m \prod_{i=1}^b\left(\epsilon^-\epsilon^* (p_i n)^2\binom{\epsilon^* p_i n +b-2}{b-2}\binom{u}{y'_{i,0}}\prod_{j=1}^b\binom{n+\epsilon^+(u-n)q_j}{y'_{i,j}}\right)
\geq \binom{u}{n}\binom{n}{p_1n,\ldots,p_bn}.\end{multline*}
Using $\frac{(a-b)^b}{b!}\leq\binom{a}{b}\leq\frac{a^b}{b!}\leq a^b$ we require
\begin{multline*}
2^m\prod_{i=1}^b\left(\epsilon^-\epsilon^*(p_i n)^b\left(1+\frac{b-2}{\epsilon^*p_in}\right)^{b}\binom{p_in}{y'_{i,0},y'_{i,1},\ldots,y'_{i,b}}u^{y'_{i,0}}\prod_{j=1}^b(\epsilon^+q_ju)^{y'_{i,j}}\left(1+\frac{n(1-\epsilon^+q_j)}{\epsilon^+q_ju}\right)^{y'_{i,j}}\right)\\\geq u^n\left(1-\frac{n}{u}\right)^n.\end{multline*}
Taking logarithms and using\[
\sum_{i=1}^b\sum_{j=0}^by'_{i,j}=\sum_{i=1}^b p_in=n\] we obtain
\begin{multline*}
m \geq -b\log(\epsilon^-\epsilon^*)-\sum_{i=1}^b b\log(p_in)-\sum_{i=1}^bb\log\left(1+\frac{b-2}{\epsilon^*p_i n}\right)-\sum_{i=1}^bp_in\H\left(\frac{y'_{i,0}}{p_in},\frac{y'_{i,1}}{p_in},\ldots,\frac{y'_{i,b}}{p_in}\right)\\+\sum_{i=1}^b\sum_{j=1}^by'_{i,j}\log(1/\epsilon^+q_j)-\sum_{i=1}^b\sum_{j=1}^by_{i,j}'\log\left(1+\frac{n(1-\epsilon^+q_j)}{\epsilon^+q_ju}\right)+n\log\left(1-\frac{n}{u}\right).\end{multline*}
Defining $r_{i,j}=y'_{i,j}/p_in$; noting that the first three terms in the previous inequality are all  $O(\log n)$ and using $\log(1+\alpha)=O(\alpha)$ for $\alpha$ small we obtain
\[
m\geq n\sum_{i=1}^bp_i\left(\sum_{j=1}^br_{i,j}\log(r_{i,j}/\epsilon^+q_j)+r_{i,0}\log r_{i,0}\right)+O(\log n)+O\left(\frac{n^2}{u}\right).\]
Dividing by $n$ and using $u\gg n$ we find that the average number of bits required per key is at least
\begin{multline}\label{av:eq}
\frac{m}{n}\geq \sum_{i=1}^bp_i\left((1-r_{i,0})\log1/\epsilon^+ + r_{i,0}\log r_{i,0}+r_{i,i}\log r_{i,i} +r_{i,i}\log 1/q_i\right)\\+\sum_{i=1}^b\sum_{j\in [b]\b\{i\}}p_ir_{i,j}\log r_{i,j}+\sum_{i=1}^b\sum_{j\in [b]\b\{i\}}p_ir_{i,j}\log1/q_j+o(1).\end{multline}
Defining $t_{i,j}=r_{i,j}/(1-r_{i,0}-r_{i,i})$ we have 
\begin{equation}\label{t:eq}
\sum_{i=1}^b\sum_{j\in [b]\b\{i\}}p_ir_{i,j}\log r_{i,j}=\sum_{i=1}^bp_i(1-r_{i,0}-r_{i,i})\left(\log(1-r_{i,0}-r_{i,i})-\H(t_{i,1},\ldots,t_{i,i-1},t_{i,i+1}\ldots,t_{i,b})\right).\end{equation}
Defining $u_{i,j}=q_j/(1-q_i)$ and applying Gibbs' inequality we obtain
\begin{eqnarray}
\nonumber \sum_{i=1}^b\sum_{j\in [b]\b\{i\}}p_ir_{i,j}\log1/q_j &\!=\!&
\sum_{i=1}^bp_i(1-r_{i,0}-r_{i,i})\left(\log 1/(1-q_i) +\sum_{j\in [b]\b\{i\}}t_{i,j}\log 1/u_{i,j}\right)\\
 \nonumber&\!\geq\! & \sum_{i=1}^bp_i(1-r_{i,0}-r_{i,i})\left(\log 1/(1-q_i) +\H(t_{i,1},\ldots,t_{i,i-1},t_{i,i+1}\ldots,t_{i,b})\right)\\
& &\label{u:eq}\end{eqnarray}
Substituting (\ref{t:eq}) and (\ref{u:eq}) into (\ref{av:eq}) yields
\begin{multline*}
\frac{m}{n}\geq \sum_{i=1}^bp_i\left((1-r_{i,0})\log1/\epsilon^+ + r_{i,0}\log r_{i,0}+r_{i,i}\log r_{i,i} +r_{i,i}\log 1/q_i\right)\\
+\sum_{i=1}^bp_i(1-r_{i,0}-r_{i,i})\left(\log 1/(1-q_i)+\log (1-r_{i,0}-r_{i,i})\right)+o(1).\end{multline*}
Defining $r_i^*=\sum_{j\in [b]\b\{i\}}r_{i,j}$ we have (by (\ref{y:eq})) that $r_i^*\leq \epsilon^*$. We also have $r_{i,0}\leq \epsilon^-$ and so $r_{i,i}=1-r_{i,0}-r_i^*\geq 1-\epsilon^- -\epsilon^*$. Hence
\begin{eqnarray}
\nonumber \frac{m}{n}&\geq& (1-\epsilon^-)\log1/\epsilon^+ -\H(\epsilon^-,\epsilon^*,1-\epsilon^--\epsilon^*)+\sum_{i=1}^bp_i\left(r_{i,i}\log1/q_i+r_i^*\log1/(1-q_i)\right)+o(1)\\
\nonumber& \geq &  (1-\epsilon^-)\log1/\epsilon^+ -\H(\epsilon^-,\epsilon^*,1-\epsilon^--\epsilon^*)+(1-\epsilon^- -\epsilon^*)\sum_{i=1}^bp_i\log1/q_i\\
\label{fi:eq}& &\hspace{8cm}\ +\ \sum_{i=1}^bp_ir_i^*\log1/(1-q_i)+o(1).\end{eqnarray}
Finally applying Gibbs' inequality and noting that the last summation in (\ref{fi:eq}) is non-negative yields our desired lower bound on the average number of bits required per key
\[
\frac{m}{n}
\geq (1-\epsilon^-)\log1/\epsilon^+ +(1-\epsilon^- -\epsilon^*)\H(\pd)-\H(\epsilon^-,\epsilon^*,1-\epsilon^--\epsilon^*)+o(1).\]
$\hfill\Box$

\medskip
\emph{Justification that $\rho$, the proportion of zeros in a Simple Bloom map,  is sharply concentrated.}

  If $Y_j$ is the expected number of bits that remain zero in the Simple Bloom map $B$, conditioned on the first $j$ hashes then $Y_0=\E[\rho m]$ while $Y_t=\rho m$. The $Y_j$ form a martingale with $|Y_{j+1}-Y_j|\leq 1$. Azuma's inequality now implies that for any $\lambda>0$ we have
\[
\P\left\{\rho<\E[\rho]-\frac{\lambda\sqrt{t}}{m}\right\}<e^{-\lambda^2/2}.\]
Hence if $t=O(m)$ then $\rho$ is extremely unlikely to be much smaller than its expected value. (Note that this argument also implies that the same is true for the more efficient Bloom maps described in Section \ref{bm:sec}.)

\medskip
\emph{Remark on $\ez$-encoding.} Having given lower bounds on the space required by $\ez$-encoding data structures in Corollary \ref{bmlb:thm} we would like to claim that the Simple Bloom map $B$ is $\ez$-encoding. This is not quite true: the \emph{expected} proportion of false positives and misassignments is at most $\epsilon$ but this does not guarantee that $B$ is $\ez$-encoding. (This is no different from the often overlooked fact that for an ordinary Bloom filter with false positive probability  $\epsilon$ the \emph{proportion} of keys in $U\b X$ for which the filter returns a false positive may be larger than $\epsilon$.)
 However the events ``$B$ returns a false positive on query $x$'', $x\in U\b X$, are independent and have probability at most $f^+(B)$. Hence if $Z$ is the number of false positives in $U\b X$ then $Z$ is stochastically dominated by  the binomially distributed variable $\tr{Bin}(u-n,f^+(B))$. Using Hoeffding's bound for the tail of the binomial distribution we have
\[
\P\{Z>(u-n)f^+(B)+\lambda\sqrt{u-n}\}\leq e^{-\lambda^2/2}.\]
Hence with high probability the proportion of false positives is at most $f^+(B)+O(1/\sqrt{u-n})$. Similarly the proportion of misassignments is (with high probability) at most $f_i^*(B)+O(1/\sqrt{n})$. Thus $B$ is essentially $\ez$-encoding. (Note that a similar argument implies that this also holds for the more efficient Bloom maps of Section \ref{bm:sec}.)

\medskip

\emph{Proof of Lemma \ref{leaf:lem}.}
First note that if $T$ is a perfect binary tree (i.e.~a full binary tree with all leaves at the same depth) then the number of nodes on $P_{i,j}$, (where $P_{i,j}$ is the part of the path from the root to $v_j$ that is disjoint from the path to $v_i$), is at least $\log (j-i+1)$. 

Now extend the tree $T$ to a tree $T'$ by replacing each leaf $v_k\in \{v_i,v_{i+1},\ldots,v_{j-1}\}$ by a perfect binary tree of depth $l_j-l_k$.  By our previous remark the number of nodes on $P_{i,j}$ is at least $\log s $, where $s$ is the number of leaves lying strictly between $v_{i-1}$ and $v_{j+1}$ in $T'$. Since $s=\sum_{k=i}^j2^{l_j-l_k}$ part (a) now follows. 

For (b) note that if we define $l_0=0$ then
\begin{eqnarray*}
\sum_{d=0}^{l_b}\frac{|T_d|}{2^d}&\leq& 1+\sum_{i=1}^b(l_i-l_{i-1})\left(1-\sum_{j=1}^i\frac{1}{2^{l_j}}\right)\\
&=& 1+\sum_{i=1}^b \frac{l_i}{2^{l_i}}.\end{eqnarray*}

For (c) note that if the path from the root to $v_i$ has left branches at depths $d_1,d_2,\ldots,d_t$ then the number of leaves to the right of $v_i$ is at least $\sum_{j=1}^t2^{l_i-(d_j+1)}$ (this is because $T$ is full). Since all of the depths of the left branches are distinct and at most $l_i-1$, the number of leaves to the right of $v_i$ is at least $\sum_{j=0}^{t-1} 2^j=2^t-1$. However the number of leaves to the right of $v_i$ is $b-i$ and so $t\leq \log(b-i+1)$. $\hfill\Box$
\end{document}